\documentclass[sigconf]{acmart}
\usepackage{multirow}
\usepackage{algorithm}
\usepackage[noend]{algpseudocode}

\usepackage{fixltx2e}

\setcopyright{none}
\usepackage{booktabs} 
\usepackage{blindtext}

\usepackage{amsmath}

\DeclareMathOperator*{\argmin}{argmin}

\title{
The effectiveness of factorization and similarity blending
}

\author{
\textbf{
Giacomo Camposampiero \enskip 
Loïc Houmard \enskip 
Marc Lundwall \enskip 
Andrea Pinto \\
}
\vspace{0.1cm}
Group: Bonjour, Department of Computer Science, ETH Zürich, Switzerland\\ 
\vspace{0.1cm}
\normalsize{\texttt{\{gcamposampie, lhoumard, mlundwall, pintoa\}@ethz.ch}}
\vspace{0.5cm}
}

\begin{document}
\begin{abstract}
{
\small
Collaborative Filtering (CF) is a widely used technique which allows to leverage past users' preferences data to identify behavioural patterns and exploit them to predict custom recommendations.
In this work, we illustrate our review of different CF techniques in the context of the Computational Intelligence Lab (CIL) CF project at ETH Zürich. After evaluating the performances of the individual models, we show that blending factorization-based and similarity-based approaches can lead to a significant error decrease (-9.4\%) on the best-performing stand-alone model.  
Moreover, we propose a novel stochastic extension of a similarity model, SCSR, which consistently reduce the asymptotic complexity of the original algorithm.
}
\end{abstract}

\maketitle

\section{Introduction}
In the last few decades, the increasing amount of information to which users are exposed on a daily basis determined the rise of recommendation systems. These models are able to identify and exploit patterns of users interest, with the goal of providing personalized recommendations and improving the final user experience. As a result, recommendation systems are now integrated in a large number of commercial applications, with some prominent examples being Amazon, Netflix, Spotify, Uber or Airbnb.

One of the most popular approaches used in the development of recommendation systems is \textit{Collaborative Filtering} (CF). In CF, past user behaviours are used to derive relationships between users and inter-dependencies among items, in order to identify new user-item associations \cite{Koren09matrixfactorization}. The main advantage of CF lies in its intrinsic domain-free nature and in its ability to model user-item interactions without requiring the creation of explicit profiles.

In this work, we explore a wide spectrum of state-of-the-art approaches to Collaborative Filtering.
In particular, we focus on both \textit{memory-based} approaches, where inference is based on calculations on past users' preference records (e.g. similarity models) and \textit{model-based} approaches, where the same past users' preferences are used to train a model which is then used at inference time to predict new recommendations (e.g. factorization, and neural models) \cite{taxonomy}. 
Other than exploring existing models, we propose a novel stochastic extension of a similarity-based algorithm, SCSR.
Finally, we empirically verify that blending factorization-based and similarity-based approaches can yield more accurate results, decreasing the validation RMSE by 9.4\% with respect to the difference between the baseline and the best-performing single model.

The performances of these model are compared on the 2022 CIL CF dataset.
This dataset consists of a sparse matrix (only $11.77\%$ of the entries are observed) that defines the interactions (integer ratings between 1 and 5) between $10000$ users and $1000$ items.

In the following sections, 
we theoretically introduce the investigated models (§\ref{sect:methods}) and explain in detail our experiments (§\ref{sect:exp}). Finally, we conclude with some final remarks on our work (§\ref{sect:conclusion}).


\section{Methods}
\label{sect:methods}
\subsection{Matrix Factorization}
As demonstrated by the Netflix Prize competition \cite{bennett2007netflix}, matrix factorization techniques are very effective in the context of CF.
The main idea behind these algorithms consists in mapping both users and items to a joint (lower-dimensional) latent factor space, such that the original interactions between users and items can be reconstructed as inner products in that latent space. 

A direct implementation of this concept can be obtained applying Singular Value Decomposition (SVD) \cite{clsvd}, a well-established technique for identifying latent semantic factors in information retrieval.
SVD decomposes the original matrix in the product
\begin{equation}
    A = U \Sigma V^T
\end{equation}
where $U\in \mathbb{R}^{m\,\times \,m}$ and $V\in \mathbb{R}^{n\,\times \, n}$ are orthogonal matrices and $\Sigma \in \mathbb{R}^{m\,\times \, n}$ is the diagonal matrix of singular values. 
In our context, matrices $U$ and $V$ could be interpreted as the latent factors associated to users and items respectively, while matrix $\Sigma$ expresses the weight associated to each latent feature. 
Since the main goal is to generalize well to unobserved entries at inference time and not obtain a null reconstruction error, the singular value diagonal matrix $\Sigma$ is approximated using only the largest $k$ singular values. Hence, the latent features for users and items are embedded in a lower $k$-dimensional space, with $k<\!\!<n,m$.

However, SVD needs a large storage space and has a large computational complexity. In order to overcome these limitations, a simple and yet remarkably powerful method, FunkSVD \cite{funk}, was proposed as part of the Netflix Prize competition. 
The algorithm initializes the latent factor matrices $U$ and $V$ randomly, and then optimize them using Stochastic Gradient Descent (SGD). The objective function optimized with SGD is:

\begin{equation}
    \argmin_{U,V} \left\lVert \, A - \widetilde{A} \,\right\rVert_F + \alpha  \left\lVert U \right\rVert + \beta \left\lVert V \right\rVert
\end{equation}

where $\left\lVert \, \cdot \, \right\rVert_F$ is the Frobenius norm, $A$ is the original matrix, and $\widetilde{A} = UV$ is the reconstructed matrix. The last two terms of the objective are regularization terms.  

Another well-known matrix-factorization technique based on iterative optimization is Alternating Least Square (ALS). The objective function which is optimized by the method is 

\begin{equation}
\label{eq:als}
    \argmin_{U,V} \left\lVert \, A - \widetilde{A} \,\right\rVert_F^2 + \lambda \left( ||U||_F^2 + ||V||_F^2 \right)
\end{equation}

Since both $U$ and $V$ are unknown, (\ref{eq:als}) is not convex. However, if we fix one of the unknowns, the optimization problem becomes quadratic and can be solved in closed form.
Thus, ALS techniques iteratively improves $U$ and $V$ by solving a least-square problem, with a loss that always decreases and, eventually, converges.

The last investigated matrix factorization approaches are Factorization Machines (FM) \cite{Rendle2010FactorizationM}, which allow for predictions over a sparse dataset. FMs have the same flexibility as Support Vector Machines (SVM) in that they accept any feature vector, yet work well over sparse data by factorizing the parameters modeling the relevant interactions. The model equation of a FM of degree 2 for $n$ features is as follows:

\footnotesize
\begin{equation}
\hat{y}(\boldsymbol{x}) := w_0 + \sum_{i=1}^n{w_i x_i} + \sum_{i=1}^n\sum_{j=i+1}^n{\langle \boldsymbol{v_i}, \boldsymbol{v_j} \rangle x_i x_j}
\end{equation}
\normalsize

\noindent
with learned parameters $w_0 \in \mathbb{R}, w \in \mathbb{R}^n$, and $V \in \mathbb{R}^{n \times k}$. $w_0$ is the global bias, $w$ models the weights of the individual features, $V$ a factorization of the weights of the interactions between pairs of features, and $v_i \in \mathbb{R}^k$ is the $i$-th row of $V$.


Adding Bayesian inference, as in Bayesian Factorization Machines (BFM) \cite{Freudenthaler2011BayesianFM}, is a considerable improvement: it allows to avoid the manual tuning of multiple hyper-parameters, and performs better than SGD and ALS optimization. 
BFM uses Gibbs sampling, a Markov Chain Monte Carlo (MCMC) algorithm for posterior inference.
with a complexity of $\mathcal{O}(k N_z)$ for each sampling step ($N_z$ being the number of non-zero elements in the feature matrix).

One of the main advantages of FM, compared to the other matrix factorization models, is that we can easily choose how to engineer the feature vectors. 
The standard way is to simply concatenate a one-hot encoding of the user with a one-hot encoding of the item: $(u, i)$, to which a list of ratings is associated.
These user-movie-rating combinations are also referred to as \textit{explicit feedback} \cite{Oard1998ImplicitFF}, i.e.  information directly provided by the users themselves. However, \textit{implicit feedback} \cite{Oard1998ImplicitFF}, which is information not directly provided by the users but that collected through their usage of the service \cite{Oard1998ImplicitFF}, is also useful to the model. For example, the date of the rating (in days since the first rating of the dataset or since the movie release) has been used successfully for this task \cite{Koren2009CollaborativeFW}. 

Even if our data does not include any implicit feedback, we can recreate some about the users by highlighting all the movies that each user has rated, resulting in the feature vector $(u, i, iu)$, as in SVD++ model \cite{Koren2008FactorizationMT}, $iu$ standing for implicit user (feedback). 
Similarly, we can also go in the other direction by showing which users have rated each movie, such as in the Bayesian timeSVD++ flipped model \cite{onthediffic}, only without the time information: $(u, i, iu, ii)$, $ii$ standing for implicit item (feedback).

\subsection{Similarity Based}
Similarity based methods are designed to find neighborhoods of similar users/items using a similarity function on common ratings. 
Within these neighborhoods, it is possible to compute weighted averages of the observed ratings (with weights being proportional to the similarity degree), and use them to infer missing entries.
In our work, we investigate 3 different similarity measures: cosine similarity, PCC and SiGra \cite{wu2017sigra}. The similarity between two users or items $u$ and $v$ are defined, respectively, as:

\footnotesize
\begin{equation*}
    \textrm{\normalsize cosine(u, v)} = \frac{\sum_{k \in I_u \cap I_v} r_{uk} \cdot r_{vk}}{\sqrt{\sum_{k \in I_u \cap I_v} r_{uk}^{2}} \cdot \sqrt{\sum_{k \in I_u \cap I_v} r_{vk}^{2}}}
\end{equation*}
\begin{equation*}
    \textrm{\normalsize PCC(u, v)} = \frac{\sum_{k \in I_u \cap I_v} (r_{uk}-\overline{\mu_u}) \cdot (r_{vk}-\overline{\mu_v})}{\sqrt{\sum_{k \in I_u \cap I_v} (r_{uk}-\overline{\mu_u})^{2}} \cdot \sqrt{\sum_{k \in I_u \cap I_v} (r_{vk}-\overline{\mu_v})^{2}}}
\end{equation*}
\begin{equation*}
    \textrm{\normalsize SiGra(u, v)} = \left(1+\exp{\left(-\frac{|I_u|+|I_v|}{2\cdot |I_u \cap I_v|}\right)}\right)^{-1}
    \cdot
    \frac{\sum_{k \in I_u \cap I_v}\frac{\min(r_{uk}, r_{vk})}{\max(r_{uk}, r_{vk})}}{|I_u \cap I_v|}
\end{equation*}

\vspace*{0.2cm}
\normalsize
\noindent
where $k \in I_u \cap I_v$ represents the indexes of all commonly rated items, respectively users and $r_{uk}$ is the rating given by user $u$ for item $k$.
These 3 functions could be either applied between all items or between all users, and we experiment with both. 

It can be observed that PCC is closely related to cosine similarity, except that it first centers the ratings by removing the average and hence try to overcome the bias which occur when some users give overall better ratings than others. 
Both PCC and cosine similarity tend to overestimate the similarity between users or items which have only few commonly rated items. 
To overcome that, some weighting functions can be added on top of the previous functions to penalize more the similarity of users or items having only few common ratings. 
We experiment with normal, significance and sigmoid weighting, defined as follows:

\footnotesize
\begin{equation*}
    {\text{\normalsize w\textsubscript{normal}(u, v)}} = \frac{2 \cdot |I_u \cap I_v|}{|I_u| + |I_v|}  \qquad {\text{\normalsize w\textsubscript{significance}(u, v)}} = \frac{\min(|I_u \cap I_v|, \beta)}{\beta} 
\end{equation*}
\begin{equation*}
   {\text{\normalsize w\textsubscript{sigmoid}(u, v)}} = \left(1+\exp{-\frac{|I_u \cap I_v|}{2}}\right)^{-1}
\end{equation*}

\vspace*{0.1cm}
\normalsize
\noindent
SiGra already have such weighting incorporated into its formula, hence we do not add any of them to it. 
Note that the significance weighting only penalize the similarity of items or users which have fewer common ratings than $\beta$ which is an hyper-parameter.

To find the neighborhoods and compute the predictions for the missing values once the similarity matrix has been computed, different methods exists. In our work, we use a weighted average of the $k$ nearest neighbors. The prediction is therefore defined as follows:

\small
\begin{equation*}
    \hat{r}_{ui} = \overline{\mu_u} + \frac{\sum_{v\in N_k^{u,i}} \textrm{sim}(u, v) \cdot (r_{vi} - \overline{\mu_v})}{\sum_{v\in N_k^{u,i}} |\textrm{sim}(u, v)|}
\end{equation*}
\normalsize
where $N_k^{u,i}$ represents the set of neighbors i.e. the set of the $k$ most similar users which have a rating for item $i$.
We also experiment with a combination of both user and item similarity for the final prediction, taking into account the confidence of the rating.

Finally, we re-implement the Comprehensive Similarity Reinforcement (CSR) algorithm \cite{hu2017mitigating}, which alternatively optimize the user similarity matrix using the item similarity matrix and vice versa. 
Note however that this method is extremely slow, as it runs in $\mathcal{O}(|I|^2 \cdot |U|^2 \cdot \textrm{max\_iter})$, and hence not applicable to our problem. 

For this reason, we propose a novel implementation of the algorithm, Stochastic CSR (Algorithm \ref{alg:scsr}) which only uses a random sample of items and users at each iteration to update the other matrix. 
This implementation of the algorithm runs in  $\mathcal{O}((|I|^2 + |U|^2) \cdot \textrm{sample\_size}^2 \cdot \textrm{max\_iter})$, which can be significantly smaller if sample\_size is small enough. 

\subsection{Neural Based}
Traditional matrix factorization techniques are often interpreted as a form of dimensionality reduction, therefore some authors investigated the use of autoencoders to tackle CF.
In this case, the input $x$ is a sparse vector and the AE's output $y = f(x)$ is its dense correspondence, containing all the rating predictions of items in the corpus. First notable AE CF model was I-AutoRec (item-based) and U-AutoRec (user-based) \cite{autorec}. 

DeepRec \cite{deeprec} is a fine-tuned version of the AutoRec model. It employs deeper networks, embraces SELU as activations \cite{selu} and apply heavy regularization techniques to avoid overfitting. 
It also proposes a variation in training algorithm to address the fixed point constraint, intrinsic to the AE CF problem objective \cite{deeprec}. 
DeepRec augments every optimization step with an iterative dense re-feeding step, i.e. it performs a classic weight update on the sparse-to-dense loss $L(x, f(x))$ and then treat $y = f(x)$ as a new artificial input, performing another weight update on the dense-to-dense loss $L(y,f(y))$.
DeepRec requires no prior training of the layers.
The model is optimized using a masked MSE loss.
During inference, model prediction is defined as a single forward pass $\hat{y} = f(x)$.

In the alternative Graph Convolutional Networks (GCN) \cite{gcn} paradigm, LightGCN (LGC) \cite{lightgcn} explores a simplified and optimized version of NGCF \cite{ngcf}, the former state-of-the-art model in GCN for CF. LGC gets rid of the feature transformation matrices and the non-linear activation function, two legacy layer propagation operations that NGCF inherited from GCN, which were argued to be not helpful for CF purposes \cite{lightgcn}. Hence, LGC propagation rule is simply defined as:

\footnotesize
\begin{align*}
e_u^{(k+1)} = \sum_{i \in \mathcal{N}_u} \frac{1}{\sqrt{|\mathcal{N}_u||\mathcal{N}_i}|} e_i^{(k)} 
\qquad\quad
e_i^{(k+1)} = \sum_{u \in \mathcal{N}_i}  \frac{1}{\sqrt{|\mathcal{N}_i||\mathcal{N}_u}|} e_u^{(k)}
\end{align*}
\normalsize

where $e_u^{(k)}$ and $e_i^{(k)}$ respectively denote the refined embedding of user $u$ and item $i$ after $k$ layers propagation, and $\mathcal{N}_u$ denotes the set of items that are interacted by user $u$ (respectively for $\mathcal{N}_i$).
The embeddings are then combined with a weighted linear combination, setting an importance to each of the $k$-th layer embeddings.
The only other trainable parameters of LGC are the embeddings of the $0$-th layer.
Originally, the model is optimized using the Bayesian Personalized Ranking (BPR) loss, but we adapt it to use a standard RMSE loss.
During inference, model prediction is defined as the inner product of user and item final representations $\hat{
y}_{ui} = e_u^T e_i$.

\subsection{Blending}
Ensemble methods are designed to boost predictive accuracy by blending the predictions of multiple machine learning models. 
These methods have been proven effective in many fields, and CF is no exception. 
For example, both the first two classified solutions in the Netflix Prize challenge exploited blending \cite{blending1, blending2}.

We propose an evaluation of different blending techniques, based on a wide range of regression approaches (linear, neural, and boosting). 
The underlying idea is to learn a transformation $\varphi:\mathbb{R}^{n\times m}\times ... \times \mathbb{R}^{n\times m} \rightarrow \mathbb{R}^{n\times m}$
such that the final prediction $\hat{A} = \varphi(\hat{A}^1,\, \dots \,, \hat{A}^k)$ is obtained in function of the predictions of $k$ different models.

\section{Experiments}
\label{sect:exp}
\subsection{Matrix factorization}
The first matrix factorization model to be implemented was ALS. 
Matrices $U$ and $V$ were initialized with the SVD decomposition of the normalized initial matrix (column-level) imputed with 0s.
The results for this model are included in Table \ref{table:results}.
A non-greedy tuning was performed, but none of the alternative parameters (rank, epoch, initialization, and imputation) improved the baseline score. 

The SVD-based factorization model was also implemented from scratch. The normalization and imputation steps were the same as for ALS. The optimal rank ($k=5$) was obtained using validation. 
A wide range of different parameters was tested, but with no improvement on the validation score. 
This model did not perform well, as shown by the validation score in in Table \ref{table:results}.

FunkSVD was on the other hand not developed from scratch. Instead, we adapted to our framework an existing Python implementation \cite{bolmier}. 
FunkSVD achieved better results compared to SVD, but did not improve on the baseline score

We used myFM Python library \cite{myFM} to implement the BFM.
Firstly, we tried the most basic version the BFM: only the one-hot encoded users and movies relevant to each rating are provided in the sparse feature matrix $(u, i)$. 
The only hyper-parameters to tune are the embedding dimension, i.e. the latent feature dimension $k$, and the number of iterations. Increasing the former decreases the validation error, reaching a plateau at around 50. Increasing the latter
generally improves the model, but the validation RMSE start decreasing beyond 500 iterations.  Therefore, for the rest of our experimentation we chose to set them at 50 and 500, respectively.

\begin{figure}[!b]
    \centering
    \includegraphics[width=.9\columnwidth]{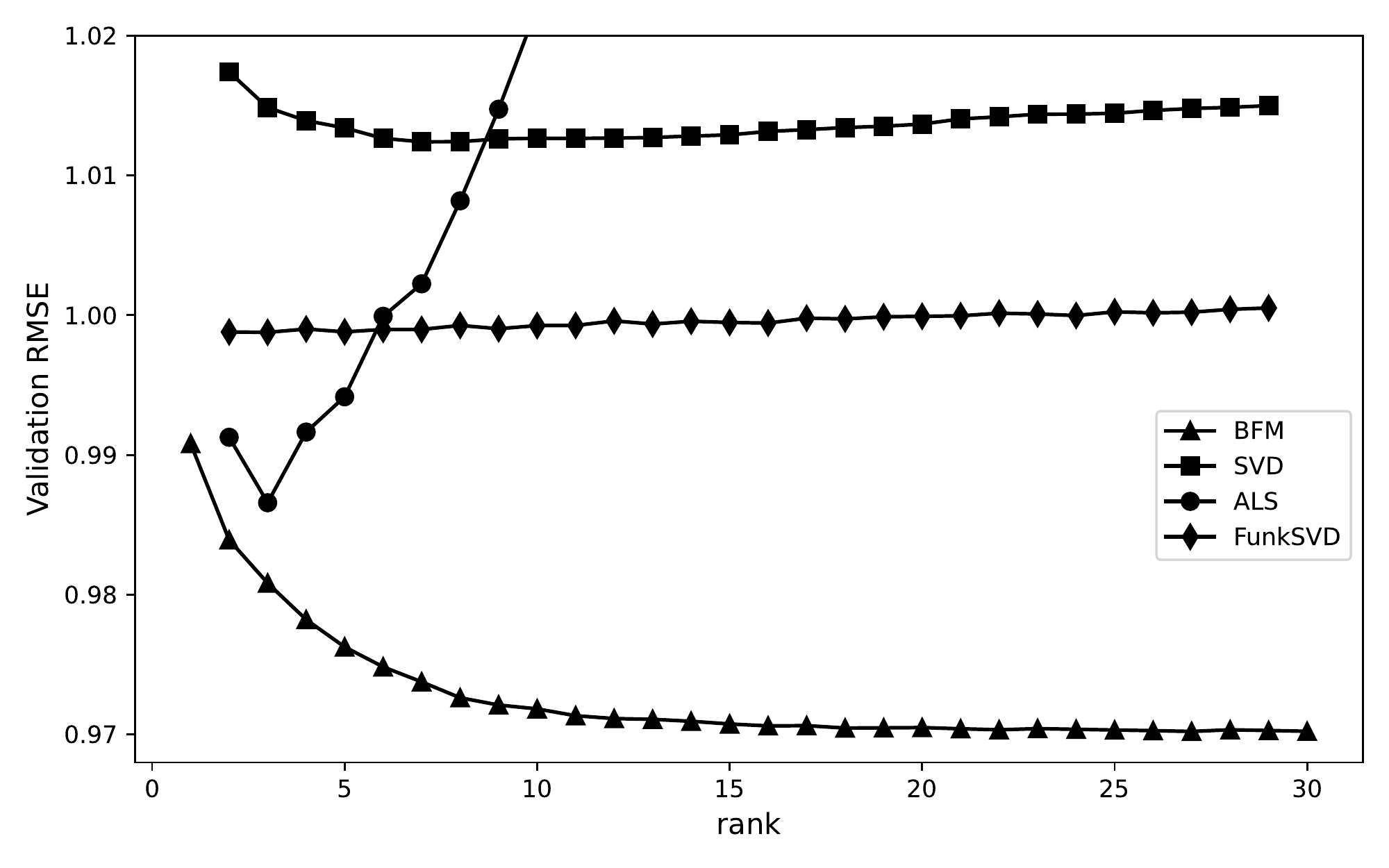}
    \caption{Rank analysis for matrix factorization techniques.}
    \label{fig:rankanalysis}
\end{figure}

We decided to experiment with adding implicit features \cite{Oard1998ImplicitFF}, as shown in the myFM \cite{myFM} documentation. 
Each row has the format $(u, i, iu)$, where $iu$, a sparse vector of size the number of movies overall, represents all the movies rated by that row's user. This is done by setting the values at the relevant indices of $iu$ to a normalized value $1/\sqrt{\left|N_u\right|}$, with $N_u$ the number of movies rated by user $u$, or to $1$ if $N_u = 0$ \cite{Koren2008FactorizationMT}. We also added implicit movie information in a similar fashion, showing all the users that have watched a movie for every movie, the complete feature matrix now being of the form $(u, i, iu, ii)$ \cite{onthediffic}. As can be observed in Table \ref{table:results}, for regression BFM the biggest change came through the addition of $iu$, bringing our validation score from 0.98034 to 0.97492. Adding $ii$ to that further improves our RMSE to 0.97268.

Furthermore, myFM allows to use ordered probit, a type of ordinal classification.
This means that the model also learns four cutpoints, separating the real domain into five sections, one for each rating.
We are then able to obtain the probabilities of a rating being in each of these categories, and we get the final rating prediction by calculating the expected value of the category. Conceptually, this way of framing the problem seems more natural, since ratings are always given as round numbers. Here, the Gibbs sampler for ordered probit on the BFM uses a Metropolis-within-Gibbs scheme \cite{myFM}, since on its own, Gibbs sampling does not work for classification problems beyond binary classification. 
The ordered probit implementation consistently outperformed the standard regression, decreasing the best RMSE from 0.97268 to 0.97032 (see Table \ref{table:results}). 
This may be explained by the fact that ratings are fundamentally discrete values and not necessarily linear, and hence ordered probit might be able to better model them.

The most interesting analysis that can be performed to compare different matrix-factorization models is on their performances when varying the number of latent features $k$.
We show the result of this investigation in Figure \ref{fig:rankanalysis}. 
It can be observed that, while all the other factorization techniques achieve a lowest point in the RMSE for lower ranks ($<10$), 
BFM continue to improve their validation score even for larger $k$. 
This behaviour was observed also in \cite{onthediffic}.

\subsection{Similarity Based}

For what concerns similarity-based models, we investigated different similarity functions, weighting functions, and a few different hyper-parameters. 
It was not possible to conduct a complete grid-search over all the possible combinations of parameters due to time constraints. 
Therefore, we fixed some of them and optimized over only the most promising.
In our experiments, we combined the 3 possible uses of similarity (item-item, user-user, and both together) with the 3 similarity functions (PCC, cosine and SiGra), the 4 weightings (normal, significance, sigmoid, and no weighting) and different number of neighbors (1, 3, 6, 10, 30 and 10000), for a total of 162 trained models.
The fixed parameters were the significance threshold $\beta$ (set to 7 for user similarity, 70 for item similarity, and 20 when we used both) and the prediction weights used when both similarity methods were used (set to 0.5). 

Finally, we also tested Stochastic CSR algorithm using the original similarity model (PCC similarity with normal weighting and all the neighbors). The model was trained for 15 iterations and a sample size of 15 items.
Table \ref{table:results} shows some of the best scoring models for each categories we obtained during this experiment.

In general, the models that used item-item similarity managed to achieve lower RMSE than the one using the user-user similarity. 
Since the loss for the user-based methods was in average higher, the ensemble using both methods also scored worse than the item-based methods with the initial user weight (0.5). However, by reducing this weight to 0.06, we managed to improve the item-based methods as we expected, since more information is available to the model.

The similarity function which achieved the best results was PCC with normal weighting. SiGra performed worse than we expected and the significance weighting or the sigmoid weighting were not as effective as the normal one. Furthermore, the best number of neighbors happened to be 60 for our best scoring model.

Finally, our novel extension of CSR algorithm, SCSR, greatly reduced the computational cost of the original algorithm, but also decreased it's effectiveness, performing worse than the model used for its initialization. We speculate that this might be due to an excessively small sample size, which might not be able to retain enough information to improve the initial solution. 
However, this algorithm was not sufficiently explored due to time constraints, and a more in-depth analysis of its performance could be carried out in future works. 

\subsection{Neural Based}
Even conducting an extensive hyper-parameters search, we were not able to bring neural-based methods to stand-alone competitive results. DeepRec still showed decent performances (Table \ref{table:results}) with an unconstrained 3 layer AE architecture (512,128,1024,128,512) that could be blended in our ensemble techniques.
LightGCN showed its strongest (yet poor) results on 4 layers with embedding size of 64.
Overall, DeepRec is fast to train (according to layers architecture) and has demonstrated decent results, while LightGCN is much longer to train (1-60) and performs systematically worse. 

\subsection{Blending}
Different blending techniques were investigated in our work. Our approach consisted in fitting different CF algorithms on a 80 split of the data and predicting the remaining 20, which became the training dataset for blending models.
We experimented with different regression models, namely (regularized) linear regression, random forest, MLP, and XGBoost. 
Inspired by \cite{Freudenthaler_bayesianfactorization}, we included in our ensemble all the BFM models, and then added other explored CF algorithms based on repeated k-fold cross-validation results. 

Table \ref{table:results} shows the results obtained by the different combinations of CF and regression models. 
Linear regression proved to be the most effective regression model (for the others we report only the RMSE on the best feature set), and our final blending model was hence of a linear combination of the different input predictions.

The most effective combination of CF models consisted of a  blending of 8 BFM and 5 similarity based models on PCC. 
An interesting phenomenon observed during this experiment was that none of the other factorization models was able to improve the ensemble score. We speculate that this happens because BFM already extract the maximum information among factorization techniques. 

On the other hand, similarity-based approaches, despite their higher RMSE, had a positive impact on the ensemble.
Blending factorization-based and similarity-based predictions allowed us to greatly decrease the validation error by an additional 9.4\% (with respect to the difference between the best single model and the baseline score), proving once again the effectiveness of blending methods in CF and showing that BFM can benefit from being combined with similarity-based approaches.

\section{Conclusion}
\label{sect:conclusion}
In this work, we presented the task of Collaborative Filtering and explored a wide range of models that can be used to tackle it. 
We showed that older matrix-factorization methods, in particular BFM, greatly outperform all the other investigated techniques, and that blending their predictions with similarity-based techniques can further improve their performances. 
Moreover, we proposed a stochastic variation of a similarity-based approach, SCSR, which consistently reduce the asymptotic complexity of the original algorithm.
Our final submission to the Kaggle competition, predicted with the ensemble of BFM and similarity models, achieved a public RMSE of 0.96374, placing our team first on the public leaderboard among the 16 teams which joined the 2022 Collaborative Filtering competition.
\clearpage

\onecolumn
\vspace*{\fill}
\renewcommand{\arraystretch}{1.2}
\begin{table*}[h]
    \centering
  \begin{tabular}{|c | c | c | l |  c  c|}
    \hline
    \textbf{Category} & \textbf{Method} & \textbf{Model} & \textbf{Parameters} & \textbf{Validation} & \textbf{Public Score} \\
    \hline
    \multirow{10}{*}{Memory-based}  & \multirow{9}{*}{Similarity} & Item (1) & PCC, normal, 30 nn. & 0.99012 & 0.98265\\
    &&Item & PCC, normal, 60 nn. &  0.98858 & 0.98174\\ 
    &&Item (2)& PCC, normal, all nn. & 0.98944 & 0.98388\\
    &&Item (3)& PCC, None, 30 nn. & 0.99069 & 0.98279\\
    &&Item (4)& PCC, None, all nn. & 0.99105 & 0.96454\\
    &&Item (6) & SiGra, all nn. & 1.00258 & -\\
    &&User & PCC, normal, all nn. & 1.00025 & -\\
    &&Both (7) & Cosine , normal, 30 nn., 0.5 w. & 0.99568 & 0.99052\\
    &&Both & PCC, normal, all nn., 0.5 w. & 0.99941 & - \\
    &&Both (5) & PCC, normal, 30 nn., 0.06 w. & 0.98767 & 0.98009\\
    &&Both & PCC, normal, 60 nn., 0.06 w. & 0.98755 & 0.98069\\
    \cline{2-6}
    &\multirow{2}{*}{Iterative similarity}  &  \multirow{2}{*}{Stochastic CSR} & PCC, normal, all nn., 0.5 w. & \multirow{2}{*}{1.00578}&\multirow{2}{*}{-}\\
    &&& 15 samples, 15 iter, $\alpha=0.5$ && \\
    \hline
    \multirow{22}{*}{Model-based}  & \multirow{10}{*}{Matrix factorization} & ALS & rank 3, $\lambda=0.1$, 20 iter  & 0.98865 & 0.98747 \\
    && SVD & rank 5 & 1.01240 & -  \\
    && FunkSVD & rank 3, $\eta$ = 1e-3, $\lambda$=5e-3 & 0.99880 & 0.99892 \\
    \cline{3-6}
    && BFM regression (8)& k = 50, 500 iters & 0.98034 & - \\
    && BFM r. [u, i, iu] (9)& k = 50, 500 iters & 0.97492 & - \\
    && BFM r. [u, i, ii] (10) & k = 50, 500 iters & 0.97773 & - \\
    && BFM r. [u, i, iu, ii] (11)& k = 50, 500 iters & 0.97268 & - \\
    && BFM ordered probit (12) & k = 50, 500 iters & 0.97668 & - \\
    && BFM o.p. [u, i, iu] (13)& k = 50, 500 iters & 0.97191 & - \\
    && BFM o.p. [u, i, ii] (14)& k = 50, 500 iters & 0.97527 & - \\
    && BFM o.p. [u, i, iu, ii] (15) & k = 50, 500 iters & 0.97032 & 0.96543 \\
    \cline{2-6}
    & \multirow{2}{*}{Neural} & DeepRec (16) & $\eta=1e-3$, batch 128, 300 epochs  &  0.98777 & 0.98559\\
    & & LightGCN & $\eta=1e-4$, batch 2048, 250 epochs &  0.99987 & 0.99929\\
    \cline{2-6}
    &\multirow{12}{*}{Blending} &  \multirow{7}{*}{Linear Regression} & (5) + (15) &  0.96968 & 0.96472\\
    &&& (5) + (6) + (7) + (15) & 0.96956 & -\\
    &&& (15) + (1)-(5) & 0.96939 & 0.96454\\
    &&& (5) + (15) + (16) & 0.969691 & -\\
    &&& (15) + (1)-(7) & 0.96940 & -\\
     &&& (8)-(15) & 0.96981 & -\\
     &&& (8)-(15) + (1)-(5) &\textbf{ 0.96870} & \textbf{0.96374}\\
    \cline{3-6}
    & &  Lasso & best, $\alpha=0.001$ &  0.96930	 & - \\
    & &  Ridge & best, $\alpha=0.01$ &  0.96870	 & - \\
    & &  MLP & best, hidden\_size=100, iter=1000 &   0.97937 & -\\
    & &  XGBoost &best,  n\_est=100, depth=7 &  0.97032 & -\\
    & &  Random Forest &best,  n\_est=100, depth=2 & 0.98367 & -\\
    \hline
  \end{tabular}
  \vspace*{0.2cm}
  \caption{Validation and submissions results.}
  \label{table:results}
\end{table*}
\vspace*{\fill}
\clearpage
\twocolumn

\bibliographystyle{IEEEtran}
\bibliography{bibliography}
\nocite{*}

\newpage
\appendix
\noindent
{\huge \textbf{APPENDIX}}
\section{Stochastic CSR pseudocode}
In the following lines, we include the pseudo-code for our novel expansion of CSR algorithm, Stochastic CSR. The notation used strictly matches the notation of the original paper.
Hence, for a more detailed explanation of the different parameters and symbols, refers to \cite{hu2017mitigating}.

\small
\begin{algorithm}
\caption{Stochastic CSR algorithm.}
\label{alg:scsr}
\small
\begin{flushleft}
\textbf{Input:} user-item matrix $R\in \mathbb{R}^{n\times m}$, damping factor $\alpha, \, \text{max\_iter}$, error threshold $\epsilon$, sampling size $\sigma$
\end{flushleft}
\begin{algorithmic}[1]
\State $U^0, V^0 \gets U_{sim}, V_{sim}$\vskip 3pt
\For{$i \in \{1, \dots, \text{max\_iter}\}$}\vskip 3pt
    \For{$a, b \in \{1\dots n\} \times \{a+1\dots n\}$}\vskip 3pt
        \State \textbf{randomly} select  $\mathcal{I}_{sa} \subset \mathcal{I}_a$,  $|\mathcal{I}_{sa}| = \sigma$\vskip 3pt
        \State \textbf{randomly} select  $\mathcal{I}_{sb} \subset \mathcal{I}_b$, $|\mathcal{I}_{sb}| = \sigma$\vskip 3pt
        \State $\mathcal{I}_S \gets \mathcal{I}_{sa} \times \mathcal{I}_{sb}$
        \State $U^i[a,b] \gets (1-\alpha) \cdot  U^{i-1}[a,b]+ \alpha \cdot \frac{\sum\limits_{i^a_{p}, i^b_{q} \in \mathcal{I}_S} w_{pq} V^{i-1}\left[i^a_p, i^b_q\right]}{\sum\limits_{i^a_{p}, i^b_{q} \in \mathcal{I}_S} \left| w_{pq} \right| }$
    \EndFor
    \For{$k, l \in \{1\dots m\} \times \{k+1\dots m\}$}\vskip 3pt
        \State \textbf{randomly} select  $\mathcal{U}_{sk} \subset \mathcal{U}_k$,  $|\mathcal{U}_{sk}| = \sigma$\vskip 3pt
        \State \textbf{randomly} select  $\mathcal{U}_{sl} \subset \mathcal{U}_l$, $|\mathcal{U}_{sl}| = \sigma$\vskip 3pt
        \State $\mathcal{U}_S \gets \mathcal{U}_{sk} \times \mathcal{U}_{sl}$
        \State $V^i[k,l] \gets (1-\alpha) \cdot  V^{i-1}[k,l]+ \alpha \cdot \frac{\sum\limits_{u^k_{c}, u^l_{d} \in \mathcal{U}_S} w_{cd} U^{i-1}\left[u_c^k, u_d^l\right]}{\sum\limits_{u^k_{c}, u^l_{d} \in \mathcal{U}_S} \left| w_{cd}\right|}$
    \EndFor
    \If{$\left\lVert \, U^i - U^{i-1} \,\right\rVert_F < \epsilon \text{ and } \left\lVert \, V^i - V^{i-1} \,\right\rVert_F < \epsilon$}\vskip 3pt
    \State \textbf{break}
    \EndIf\vskip 3pt
\EndFor
\State \textbf{return} $U^i, V^i$
\end{algorithmic}
\end{algorithm}

\normalsize
\begin{flushleft}
Where $w_{cd}$ (respectively $w_{pq}$) is defined as in the original paper
\begin{align*}
    w_{cd} = 1-2|r_{u^{k}_{c}k} - r_{u^{l}_{d}l}|\\
    w_{pq} = 1-2|r_{i^{a}_{p}a} - r_{i^{b}_{q}b}|
\end{align*}
where $r_{u^{k}_{c}k}$ is the normalised rating value given by user $u^{k}_{c}$ to item $i_k$ and $r_{u^{l}_{d}l}$ is that given by user $u^{l}_{d}$ to item $i_l$ (same logic for $r_{i^{a}_{p}a}$ and $r_{i^{b}_{q}b}$).
\end{flushleft}

\end{document}